\newcommand{\be}{\begin{equation}}
\newcommand{\ee}{\end{equation}}
\newcommand{\bea}{\begin{eqnarray}}
\newcommand{\eea}{\end{eqnarray}}
\newcommand{\ba}[1]{\begin{array}{#1}}
\newcommand{\ea}{\end{array}}

\documentclass[twocolumn,aps,showpacs,preprintnumbers,amsmath,amssymb]{revtex4}
\usepackage{epsfig}
\usepackage{amssymb}

\begin{document}

\title{Entanglement of two
distant Bose-Einstein condensates by detection of Bragg-scattered photons}

\author{B. Deb}

\address{Department of Materials Science,  and Raman Center for
Atomic, Molecular and Optical Sciences, Indian Association for the Cultivation of Science, Jadavpur, Kolkata
- 700032.}

\author{G. S. Agarwal}

\affiliation{Department of Physics, Oklahoma State University, Stillwater, Oklahoma 74078, USA}

\date{\today}

\begin{abstract}
We show that it is possible to generate entanglement  between two distant  Bose-Einstein condensates by
detection of Hanbury Brown-Twiss  type correlations in photons Bragg-scattered by the condensates. Upon
coincident detection of two photons by two detectors, the projected joint state of two condensates is
shown to be  non-Gaussian.  We verify the existence of  entanglement by showing that the partially  transposed state is negative. Further we use the inequality in terms of higher order moments to confirm entanglement. Our proposed scheme can be generalized for multiple
condensates and also for spinor condensates with Bragg scattering of polarized light with the latter capable
of producing hyper entanglement.
\end{abstract}

\pacs{PACS numbers: 03.75.Fi,03.65.Ud,32.80.-t,42.50.Dv} \maketitle

\section{introduction}

Quantum entanglement means inseparability of joint wave function of two or more distant objects into a
product of wave functions of individual objects -
 even in the absence of any mutual interaction or
communication between them. This epitomizes the underlying nonlocal character of quantum world. One of the
consequences of this nonlocal realism is that a single local measurement can not reveal the complete state
of an entangled system, since the process of measurement itself forces  the wave function to ``collapse"
into one its measured (eigenstate) state in a probabilistic sense. Thus, measurement process can
post-selectively play a role in creation and manipulation of entanglement, and this is the essence of what
is called ``projective measurement".

Efficient generation of entanglement in  many-particle systems and its robust transmission and transfer to
other systems is important for quantum information processes. Based on atom-photon interaction and the
exchange of photons between the qubits, entanglement in distant atomic
states~\cite{Julsgaard:2001:a,Matsukevich:2006:a,Chou:2005:a} and also between  photons
~\cite{Bouwmeester:1999:a,Kiesel:2006:a} have been experimentally demonstrated. There is another way of
entangling two remote systems without requiring any direct interaction between them: This is based on
projective measurement. This indirect method of creating entanglement between distant systems can also be
applied for many quantum communication tasks. In a recent experiment, Moehring {\it et al.} \cite{monroe}
have created entanglement between two distant trapped ions by coincident detection of two photons
spontaneously emitted by the two ions. An earlier experiment has shown interference of light emitted by two
atoms~\cite{Beugnon:2006:a,Maunz:2007:a,Maunz:2007:b} making  use of projective measurements. Thiel {\it et
al.} \cite{thiel} have proposed a scheme of entangling several remote atomic qubits and thereby  creating
Dicke state~\cite{Dicke:1954:a} of many-atoms by projective measurement of photons using multiple
photo-detectors. Dicke states are particularly important for their  robustness against particle
loss~\cite{Stockton:2003:a,Bourennane:2006:a} and non-local properties of  entangled multipartite
states~\cite{Toth:2005:a,Usha:2007:a,Retzker:2007:a,Kiesel:2006:a}. There are several other proposals for
projecting distant non-interacting particles into entangled states via
photo-detection~\cite{Cabrillo:1999:a,Bose:1999:a,Skornia:2001:a,Simon:2003:a,Duan:2003:a,Duan:2001:a}.
Continuous variables like the quadratures of a field  mode (which are analogous to position and momentum)
have also been employed \cite{kimble} in entanglement studies.

Bose-Einstein condensate (BEC)  is a macroscopic quantum object where entanglement arises quite naturally
due to two-body interaction. Bogoliubov theory \cite{bogoliubov} of Bose condensation reveals that in the
ground state of condensate, two particles with opposite momentum are maximally  entangled \cite{deb1} in
momentum variables as in EPR state \cite{epr}. This unique feature makes Bose condensates a good source of
entanglement in motional degrees of freedom.  Furthermore, in a two-component BEC, one can generate
entanglement in  hyperfine spin degrees of freedom \cite{sorensen,duan,pu,you,moore,raghavan,bigelow}. In
order to extract the intrinsic entanglement of a BEC for useful purpose of quantum information processes, it
is required to excite quasi-particles in momentum modes by stimulated Raman scattering or  Bragg scattering
\cite{quasi,bragg}. Then a scattered atom becomes entangled with Bragg-scattered photon \cite{gasenzer}. In
fact, Bragg spectroscopy can be used as a tool for generating entanglement of different kinds in a variety
of physical situations. For instance, tripartite entanglement among two momentum modes of BEC and one
electromagnetic field mode can be produced \cite{deb1}. Furthermore, it has been shown that when a common
laser beam passes through two spatially separated condensates, photon scattered by the first condensate
carries and transfers quantum information to the second one and thereby two condensates become entangled
\cite{deb2}. Similar experimental scheme has been used to produce and subsequently measure phase difference
between two spatially separated condensates \cite{phase}.

 \begin{figure}
 \includegraphics[width=2.0in]{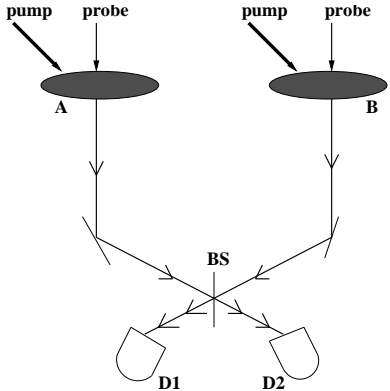}
 \caption{A scheme for entangling two separate BECs by Bragg scattering and photo-detection. A and B are two
 BECs which scatter photons from pump beams  into probe ones. $D_1$ and $D_2$ are two photo-detectors and
 `BS' stands for beam splitter. }
 \label{figdel}
 \end{figure}

Here we propose a new scheme for generation of entanglement
 between two remote
condensates by projective measurement on Bragg-scattered photons. Our scheme relies on coincident detection
of two Bragg-scattered photons coming from two remote condensates in a Hanbury Brown-Twiss type experimental
arrangement as schematically illustrated in Fig.1. A and B are two remote single-component condensates. Bragg
scattering of pump photons occur independently at the two condensates, the scattering is stimulated by probe
beams. This gives rise to the generation of quasi-particles in both the condensates in particular momentum
modes determined by the relative angle between pump and probe light beams. The probe beams are assumed to
have the same mode, that is, same frequency and polarization property. The scattered photons in the probe
modes are allowed to pass through a beam splitter in order to coalesce them to lose which path information.
Then the two photons coming out of the output ports of the beam splitter are detected by two photo-detectors
$D_1$ and $D_2$ in a coincident way. The joint projective state of the two spatially separate condensates is
explicitly non-Gaussian. The entanglement of such non-Gaussian states can  not be ascertained from the
correlation in fluctuations of quadrature phase variables  or  number variables \cite{gsabiswas}. To prove the
existence of entanglement in such a non-Gaussian state, we use the inequality criterion  recently proposed by Agarwal and
Biswas \cite{gsabiswas} based on the phase fluctuation  of paring operators (of the form  $\hat{a}\hat{b}$ ,
where $\hat{a}$ and $\hat{b}$ are annihilation operators of the two sub-systems)  and the concept of partial
transpose of Peres \cite{peres} and Horodecki \cite{horodecki}.  Vogel \cite{vogel} and also Hillery and Zubairy \cite{hillery} have recently introduced a class of similar inequalities the violation of which is sufficient to show the presence of entanglement in two bosonic  fields.

The paper is organized in the following way. Since Bragg scattering holds the key for generating
quasi-particles and atom-photon entanglement, we first give a brief introduction to Bragg scattering in
Sec.II in order to reveal the essential physical processes involved in Bragg scattering. We then formulate
our theoretical model of projective measurement and discuss its effect in generating entanglement  in
Sec.III.  In Sec.IV, we characterize entanglement and discuss our results. The paper is concluded in Sec.V.

\section{Bragg scattering}

In a Bose condensate of weakly interacting atomic gases, zero momentum  ($\mathbf{k}=0$) state is
macroscopically occupied. Therefore,  atom-atom collision occurs primarily between zero- and non-zero
momentum atoms. In stimulated Raman or Bragg scattering, two far-off resonant laser beams  with a small
frequency difference is impinged on a trapped BEC. There are basically two physical processes in Bragg
scattering. In the first process, a photon from the laser beam with higher frequency is scattered into a
photon of the other laser mode. This  causes transformation of  a zero-momentum atom  into an atom of
momentum $\mathbf{q}$, where $\mathbf{q}$ is the difference in photon momentum of the two beams.  In the
second process, an atom moving with a  momentum $-{\mathbf{q}}$ is scattered into a zero-momentum state.
Because of bosonic stimulation, the scattering of atoms from zero- to $\mathbf{q}-$ momentum state will be
dominant process. There also occur the processes which are opposite to above two processes, but these are
subdued due to phase mismatch. Thus Bragg scattering generates quasiparticles \cite{quasi}, predominantly in
two momentum side-modes $\mathbf{q}$ and $-\mathbf{q}$. Bragg spectroscopy \cite{bragg} with coherent or
classical light produces  coherent states of quasiparticles in a BEC. When these quasiparticles are
projected into particle domain, they form two-mode squeezed as well as entangled state \cite{deb1} in
particle number variables.

Two remote condensates A and B are subjected to Bragg scattering with pairs of Bragg pulses. The frequencies
and the directions of propagation of the laser beams are so chosen  such that Bragg resonance (phase
matching) conditions of scattering  in both  the condensates are fulfilled.  We assume the pulse with higher
frequency has higher intensity and hence can act as a pump. The other laser beam which acts as stimulant for
scattering is of much lower intensity and so can be considered as probe beam. We treat pump beams
classically. Since scattering at the two condensates occur independently, the Hamiltonian is simply the sum
of the Hamiltonian $H^{A}$ and $H^{B}$ corresponding to condensates A and B respectively. Let $\hat{a}_q$
and $\hat{b}_q$ represent the annihilation  operators for particles with momentum $\mathbf{q}$ in condensate
A and B, respectively. Let the corresponding Bogoliubov quasiparticles be denoted by $\hat{\alpha}_q$ and
$\hat{\beta}_q $, respectively. In terms of  these quasiparticle operators, the effective Hamiltonian as
derived in the Appendix can be written as
\begin{eqnarray}
H_{eff}^{J} = &\hbar \omega_q^B&\left(\hat{\chi}_{{\mathbf{q}}}^{\dagger} \hat{\chi}_{{\mathbf{q}}}
+\hat{\chi}_{-{\mathbf{q}}}^{\dagger} \hat{\chi}_{-{\mathbf{q}}}\right) -\hbar\delta_j\hat{c}_{j}^{\dagger}
\hat{c}_{j} \nonumber \\ &+&\left[\hbar\eta \hat{c}_{j}^{\dagger} (\hat{\chi}_{\mathbf{q}}^{\dagger} +
\hat{\chi}_{-{\mathbf{q}}}) +{\mathrm H.c.}\right]
\end{eqnarray}
where the superscript $J$ stands for condensate A or B, $\hat{c}_j$ (with $j=a,b$) is photon annihilation
operator for the probe beam  applied to condensate $J$. The particle operators $\hat{\pi}_q (\equiv
\hat{a}_q, \hat{b}_q)$ are related to the quasi-particle operators $\hat{\chi}_q (\equiv \hat{\alpha}_q,
\hat{\beta}_q )$ by Bogoliubov's transformation
\begin{eqnarray}
\hat{\pi}_{q}= u_{q}\hat{\chi}_{q}-v_{q}\hat{\chi}_{-q}^{\dagger}
\end{eqnarray}  with \begin{eqnarray} v_{q}^2 = (u_{q}^{2}-1) =
\frac{1}{2}\left (\frac{\hbar \omega_{q} + \mu}{\hbar \omega_q^B}-1 \right )\end{eqnarray}  and
\begin{eqnarray}
\hbar \omega_q^B=\left[(\hbar \omega_{q} + \mu)^2 - \mu^2\right]^{1/2}
\end{eqnarray}
is the energy of Bogoliubov's quasiparticle. Here $\hbar \omega_q = \hbar^2 q^2/(2m)$ is the kinetic energy
of a single atom, $\mu = \frac{\hbar^2\xi^{-2}}{2m}$ is the chemical potential with $\xi = (8\pi
n_{0}a_{s})^{-1/2}$ being  the healing or coherence length,  $\delta_j$ is the detuning between the pump and
probe frequencies, $\eta = \sqrt{N}f_q\Omega$, where $f_q = u_q-v_q$ and $\Omega$ is the two-photon Rabi
frequency. We assume the Bragg resonance condition ($\delta \simeq \omega_{q}$).
 The hamiltonian can
be solved exactly  in Heisenberg picture. The Heisenberg equations of motion for a triad of operators $X =
\left(\hat{\alpha}_{{\mathbf{q}}} \hspace{0.2cm} \hat{\alpha}_{-{\mathbf{q}}}^{\dagger} \hspace{0.2cm}
\hat{c}_{j}^{\dagger}\right)^{T}$ can be written in a matrix form $\dot{X} = i\omega_q^B{\mathbf{M}}X$,
where $\mathbf{M}$ is a $3 \times 3$ matrix
\begin{equation}
{\mathbf{M}} = \left(
\begin{array}{ccc}
-1 & 0 & -\tilde{\eta} \\ 0 & 1 & \tilde{\eta} \\ \tilde{\eta}^{*} & \tilde{\eta}^{*} & -\tilde{\delta}\\
\end{array}
\right)
\end{equation}
where $\tilde{x} = \tilde{x}/\omega_q^B$.  Let $\mathbf{D}$ be the diagonalizing  matrix of $\mathbf{M}$.
The solutions can be explicitly written as
\begin{eqnarray}
X(t) = {\mathbf{D}}{\mathbf{E}}(t){\mathbf{D}}^{-1}X(0) \label{soln} \end{eqnarray} where $\mathbf{E}$ is a
diagonal matrix :  ${\mathbf{E}} = \mathrm{ {\bf diag}}.[\exp(i\lambda_{1}\tau), \exp(i\lambda_{2}\tau),
\exp(i\lambda_{3}\tau)]$ with $\tau = \omega_q^B t$ and $\lambda_{i}$s being the eigenvalues of $\mathbf{M}$
matrix.

\section{Entanglement produced by photo-detection}

Our proposed scheme is shown in Fig.1. Quasiparticles are generated in the condensates A and B due to
stimulated light scattering in a pump-probe type Bragg-spectroscopic method. Let $\hat{c}_{a}$ and
$\hat{c}_b$ denote annihilation operators for the two probe light beams scattered by condensates A and B,
respectively. Using Eq. (\ref{soln}),  the scattered light at the output of the two condensates can be
represented by
\begin{eqnarray} \hat{c}_a(t) =
a_q(t) \hat{\alpha}_q^{\dagger} + a_{-q}(t)\hat{\alpha}_{-q} + a_c(t)\hat{c}_{a}(0) \label{ca}
\end{eqnarray}
\begin{eqnarray} \hat{c}_b(t) =
b_q(t) \hat{\beta}_q^{\dagger} + b_{-q}(t)\hat{\beta}_{-q} + b_c(t)\hat{c}_{b}(0) \label{cb} \end{eqnarray}
where $a_{\pm q}$, $b_{\pm q}$, $a_c$ and $b_c$ are time-dependent coefficients determined by the Eq.
(\ref{soln}). The scattered light output coming from the two condensates are passed through a beam-splitter
and finally collected at the two detectors at D1 and D2 as shown in Fig.1. Let the reflectivity and
transmissivity at left side of the beam-splitter are $r$ and $t$, respectively, while those at right side
are $r'$ and $t'$. Then the photon annihilation operators  at D1 and D2 can be expressed as
\begin{eqnarray}
\hat{C}_{D_1} = t' \hat{c}_b + r \hat{c}_a \label{cd1} \\ \nonumber \\ \hat{C}_{D_2} = t \hat{c}_a + r'
\hat{c}_b \label{cd2}
\end{eqnarray}
Let the initial state of the total system i.e. two condensates plus the two probe fields, be represented by
\begin{equation}
\mid \Psi_0 \rangle  = \mid 0,0 \rangle_{AB} \mid \alpha, \beta \rangle_{fields} \end{equation} where $\mid
0,0 \rangle_{AB}$ indicates a product state with both the condensates in ground states of quasiparticles,
where  first `0' corresponds to condensate A and the second `0' to condensate B. We assume that both the
probe fields are in coherent states $\mid \alpha, \beta \rangle_{fields}$ where the field amplitudes
$\alpha$ and $\beta$ correspond to the probes incident at A and B, respectively. Measurement of two-photon
correlation via coincident detection of scattered probe lights at the two detectors will project the
two-condensate density operator into
\begin{eqnarray}
 \rho_{AB} = {\cal N} \mathrm{Tr}_{\rm{fields}} \hat{C}_{D_2}
\hat{C}_{D_1} \rho_0 \hat{C}_{D_1}^{\dagger} \hat{C}_{D_2}^{\dagger} \label{condmeas}
\end{eqnarray}
where  $\mathrm{Tr}_{\rm{fields}}$ implies tracing over the field states, $\rho_0 = \mid \Psi_0 \rangle
\langle \Psi_0  \mid $ and ${\cal N}$ denotes a normalization factor. Now, substituting \ref{cd1} and
\ref{cd2} into \ref{condmeas} and using the relations \ref{ca} and \ref{cb}, we obtain $ \rho_{AB} = \mid
\Phi \rangle \langle \Phi \mid $ where
\begin{eqnarray}
\mid \Phi \rangle &=&  \sqrt{{\cal N}} \langle \alpha, \beta \mid \hat{C}_{D_2} \hat{C}_{D_1} \mid \Psi_0
\rangle  \nonumber \\ &=& \sqrt{{\cal N}} \left ( r't' \mid 0_A, S_B(1_q,2_q) \rangle + r t \mid
S_A(1_q,2_q), 0_B \rangle \right ) \nonumber
\\ &+&  \sqrt{{\cal N}} (r' r + t' t) \mid \Sigma_A(1_q), \Sigma_B(1_q)
\rangle. \label{reducedstate}
\end{eqnarray}
The states $\mid S_j(1_q,2_q) \rangle$ denotes a superposition state of  ground, one and two $q$-phonon
excited states of condensate $j(\equiv A, B)$. Similarly, $ \mid \Sigma_j(0, 1_q), \rangle$ is another
superposition state of ground and one phonon excited states of condensate `j'. Explicitly, these
superposition states  can be expressed as
\begin{eqnarray}
\mid S_A \rangle &=& \sqrt{2} a_{q}^2\mid 2_q \rangle_A + 2 a_q a_c \alpha \mid 1_q \rangle_A \nonumber \\
&+&  a_c^2 \alpha^2 \mid 0\rangle_A \\ \nonumber \\ \mid \Sigma_A \rangle &=& a_q \mid 1_q \rangle_A + a_c
\alpha \mid 0\rangle_A
\end{eqnarray}
 Now, we
have the reciprocity relations
\begin{eqnarray*}
r^* t' + r' t^* &=& 0, \hspace{1.0cm} r^* t + r' t'^* =0 \\
\\
|r'| = |r|, \hspace{.5cm} |t'| &=& |t|, \hspace{.5cm} |r|^2 + |t|^2 = 1
\end{eqnarray*}
Let $t = |t| \exp(i\phi)$ and $t' = |t'| \exp(i\phi')$. Since phase changes by $\pi/2$ on reflection, we
have $r = i |r| \exp(i\phi)$ and $r' = i |r'| \exp(i\phi')$. Using the reciprocity relations and considering
the field amplitudes $\alpha$ and $\beta$ as real quantities,
 we obtain

\begin{eqnarray}
\mid \Phi \rangle &=& \sqrt{{\cal N}} |r||t|\left ( \exp(2i\phi') \mid 0_A, S_B \rangle +  \exp(2i\phi) \mid
S_A, 0_B \rangle \right ) \nonumber
\\ &+&  \sqrt{{\cal N}} \exp[i(\phi + \phi')] (|r|^2 -  |t|^2) \mid \Sigma_A, \Sigma_B
\rangle. \label{reducedstate}
\end{eqnarray}
For a 50:50 beam splitter, we then have
\begin{eqnarray}
\mid \tilde{\Phi} \rangle = \sqrt{{\cal N}} \frac{1}{2} \left ( \mid 0_A, S_B \rangle +  \exp(2i\Delta \phi)
\mid S_A, 0_B \rangle \right ) \label{5050}
\end{eqnarray}
where $\Delta \phi = \phi - \phi'$ and $\mid \tilde{\Phi} \rangle
 = \exp(-2i \phi') \mid \Phi \rangle$.   Eq. (\ref{5050}) is  manifestly an entangled state of the two
 condensates A and B. This state obtained via two-photon detection is different from the Gaussian state of
 each independent Bose condensate. Furthermore, the basis states involved in this entanglement describe
 collective modes or phonon modes of the condensates, $\mid 0_A, S_B \rangle$ refers to a joint condensate
 state in which condensate A is in zero phonon and condensate B is in a superposition of zero, one and two
 phonon states.
This state results from quantum interference of two probable processes which are: (1) two scattered photons
come from condensate B and no photon is scattered by condensate A, these two photons are then split by the
beam splitter and detected at the two detectors projecting the joint condensate state into the form $\mid
0_A, S_B \rangle$. (2) The second process consists of scattering of two photons by condensate A and no
photon by condensate B, this results in the joint condensate state $\mid S_A, 0_B \rangle$. Since these two
processes are probabilistic, the quantum interference of these two processes eventually gives rise to the
resultant state of the form (\ref{5050}) for a 50:50 beam splitter. When one scattered photon comes from A
another from B, there is the probability amplitude of $r r'$ that both the detectors will detect the only
reflected part of the two photons and also the probability amplitude of $t t'$ that only transmitted part of
both the photons will be detected. Both these processes project the joint condensate state into the form $
\mid \Sigma_A(1_q), \Sigma_B(1_q) \rangle$. The net probability amplitude being the sum of these two
quantities, for a 50:50 beam splitter they cancel each other. This explains why there is no component of $
\mid \Sigma_A(1_q), \Sigma_B(1_q) \rangle$  in Eq. (\ref{5050}).

 \begin{figure}
 \includegraphics[width=3.25in]{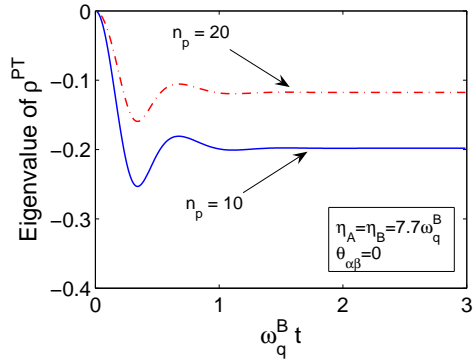}
 \caption{ One eigenvalue of the partial-transposed density matrix $\rho^{A,B^T}$ calculated in
the quasiparticle picture is plotted as a function of dimensionless interaction time
 $\omega_q^B t$,  for different values of average probe photon numbers
 $n_p = 10$ (solid), $n_p = 20$ (dashed-dotted). The phase difference is
 $\Delta \phi = 0$ and the phase difference between the two input laser fields $\theta_{\alpha\beta} = 0$.
 The effective atom-field
 coupling constants are  $\eta_B = \eta_A = 7.7 \omega_q^B$. }
 \end{figure}

As discussed earlier,  photons scattered by a condensate share entanglement  with phonons or condensate
momentum modes. These scattered photons can act as the carriers of quantum information of phonons. In a BEC
of weakly interacting atomic gases, phonons are long-lived.  By coincident detection of two independent
photons but  separately entangled with their respective scatterer condensates,
 we can establish a quantum communication channel between  two condensates . Thus our proposed method of
 generating entanglement between two remote condensates may also find application in quantum cryptography
 \cite{crypto} and teleportation \cite{teleport}.
Furthermore, this two-photon detection scheme may enable  a partial Bell state analysis \cite{bell,
bellanal,Simon:2003:a} . Most of the earlier proposals for creating entanglement between two distant ions or
atomic ensembles via photo-detection are based on the  electronic excitation and subsequent emission of
photons that are detected. In such situations, spontaneous decay to unwanted electronic states and
decoherence can not be avoided. In our proposed scheme, since we use far-off resonant stimulated Raman-type
light scattering, spontaneous emission is negligible. Since the system is a Bose condensate of weakly
interacting ultracold atoms with long coherence time, decoherence is also at a minimum level.   Moreover, we
use coherent light for stimulating photon scattering and so our scheme does not require any cavity.

\section{Characterizing entanglement in state (17)}
\subsection{Negativity of the partial transpose of the density matrix}
The necessary and sufficient condition for entanglement in any bipartite system is the negativity of at least
one of the eigenvalues of the partial transpose \cite{peres} of the density matrix of the system. Let
$\rho^{A,B}$ denote the density matrix of a bipartite system composed of sub-systems A and B.  Under partial
transpose of Peres and Horodecki over the sub-system B (or A), let the density matrix be  represented by
$\rho^{A,B^T}$  (or $\rho^{A^T,B}$ ) which can be derived by making transpose only on the operators of  B
(or A) . If the wavefunction of the composite system is inseparable, then there will be at least one
eigenvalue of the transposed  matrix which is negative.

Now, the state in Eq.(17) can be expressed as \bea \mid \tilde{\Phi} \rangle = C_0 \mid 0, 0 \rangle +
\sum_{m=1}^{2} \left [  \mid C_{m, 0} \mid m, 0 \rangle +   \mid C_{ 0 m} \mid 0, m \rangle \right ]
\label{newstate} \nonumber \\ \eea where $\mid m, n \rangle $ represents a joint quasiparticle number basis with $m$
number of quasiparticles in condensate A and  $n$ number of quasiparticles in condensate B. From Eqs. (14)
and (17), we find \bea C_0 = a_c^2 \alpha^2 + \exp(2 i \Delta \phi) b_c^2 \beta^2.  \label{c0} \eea
Similarly, all other coefficients $C_{m,n}$ can be deduced from Eqs. (14) and (17). The density operator  is
$\rho^{AB} = \mid \tilde{\Phi} \rangle \langle \tilde{\Phi} \mid$. Taking partial transpose on $\rho^{AB}$
with respect to B implies changing the base operators $\mid i j \rangle \langle m n \mid \rightarrow \mid i
n \rangle \langle m j \mid$ and the resulting matrix is $\rho^{AB^T}$.   We numerically find that in all
parameter regimes there is one eigenvalue of $\rho^{AB^T}$ which is always negative and hence the state (17)
is an entangled state. Figure 2 shows that the negativity of the eigenvalue is more prominent when the
probe photon number $|\alpha|^2 = |\beta|^2 = n_p$ is lower.

\subsection{Violation of the entanglement inequalities for observables}
Based on the idea of partial transpose, Simon  and also Duan {\it et al.} \cite{zollersimon} have
independently given an entanglement criterion.
 Using variances in
quadrature variables, an entanglement parameter  \cite{zollersimon} can be defined as \bea \xi_{XP} =
\frac{1}{2} \left [ \langle \Delta (\hat{X}_{A} + \hat{X}_{B})^2 \rangle + \langle \Delta (\hat{P}_{A} -
\hat{P}_{B})^2 \rangle \right ]. \eea where $\hat{X}_{S\equiv A,B} = (1/\sqrt{2})(\hat{S} +
\hat{S}^{\dagger})$ and $\hat{P}_{S} = (1/\sqrt{2}i)(\hat{S} - \hat{S}^{\dagger})$  with $\hat{S}$ being
any bosonic operator of the sub-system $S$. According to the criterion of \cite{zollersimon},
the condition for the occurrence of entanglement is $\xi_{XP} <
1$. This condition is necessary and sufficient for Gaussian quadrature variables only.   For non-Gaussian states, this is only sufficient. There are certain non-Gaussian bipartite
states which are conspicuously inseparable but do not  fulfill the condition $\xi_{XP} < 1$. In the present context, the state (17) is manifestly a non-Gaussian state. This prompts us to look for other criteria \cite{gsabiswas,vogel,hillery}
based on higher order moments of
observables.
\begin{figure}
 \includegraphics[width=4.125in]{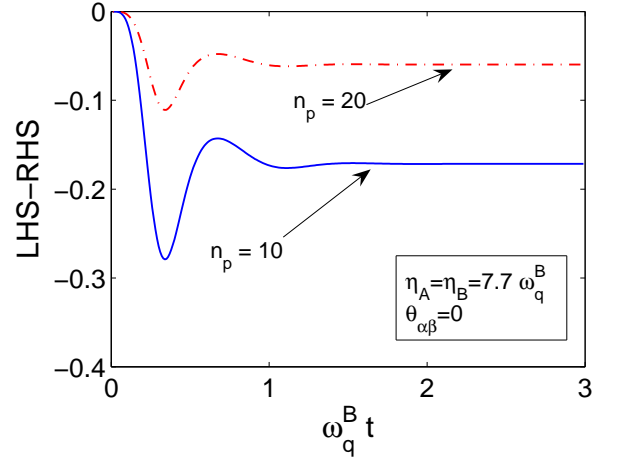}
 \caption{Left hand side (LHS) minus right hand side (RHS) of the entanglement inequality (\ref{ineq})
 calculated
in the quasiparticle picture is plotted as a function of  $\omega_q^{B} t$ for  average probe photon numbers
$n_p  = 10$ (solid) and $n_p = 20$ (dashed) . All other
 parameters are same as in Fig.2. }
 \label{figdel}
 \end{figure}
Let us now test whether the higher order entanglement criterion  introduced in Ref. \cite{gsabiswas}  can
reveal the entanglement in state (17). To this end,  let us first discuss what this criterion is. Using quasiparticle operators $\hat{\alpha} $ and $\hat{\beta}$ (same as $\hat{\alpha}_q $ and $\hat{\beta}_q$ defined in Eq.(2),  the subscript $q$ is omitted for simplicity), we construct the following operators \bea
K_x = \frac{1}{2} \left ( \hat{\alpha}^{\dagger} \hat{\beta}^{\dagger} + \hat{\alpha}\hat{\beta} \right ) \eea \bea K_y =
\frac{1}{2i} \left ( \hat{\alpha}^{\dagger} \hat{\beta}^{\dagger} - \hat{\alpha}\hat{\beta} \right ) \eea \bea K_z=
\frac{1}{2} \left ( \hat{\alpha}^{\dagger} \hat{\alpha} + \hat{\beta}^{\dagger}\hat{\beta} + 1 \right ) \eea which satisfy
SU(1,1) algebra. These SU(1,1) operators are previously employed for studying higher order squeezing. The
Heisenberg uncertainty relation of these operators implies the inequality \bea \Delta \left (
\hat{\alpha}^{\dagger} \hat{\beta}^{\dagger} + \hat{\alpha}\hat{\beta} \right )
 \Delta  \left( \frac{ \hat{\alpha}^{\dagger} \hat{\beta}^{\dagger} - \hat{\alpha}\hat{\beta}}{i} \right )
\ge \left \langle  \hat{\alpha}^{\dagger} \hat{\alpha} + \hat{\beta}\hat{\beta}^{\dagger} \right \rangle \nonumber \\ \eea Let
us now make partial transpose $\hat{\beta} \leftrightarrow \hat{\beta}^{\dagger}$. Under this partial transpose ,
the above inequality becomes \bea \Delta \left ( \hat{\alpha}^{\dagger} \hat{\beta}+ \hat{\alpha}\hat{\beta}^{\dagger} \right
)
 \Delta  \left( \frac{ \hat{\alpha}^{\dagger} \hat{\beta} - \hat{\alpha}\hat{\beta}^{\dagger}}{i} \right )
\ge \left \langle  \hat{\alpha}^{\dagger} \hat{\alpha} + \hat{\beta}\hat{\beta}^{\dagger} \right \rangle \label{ineq1}
\nonumber \\ \eea After some simple algebra, we have \bea \Delta \left ( \hat{\alpha}^{\dagger} \hat{\beta}+
\hat{\alpha}\hat{\beta}^{\dagger} \right )^2 =  N_2 +  N + M \eea \bea \Delta  \left( \frac{ \hat{\alpha}^{\dagger}
\hat{\beta} - \hat{\alpha}\hat{\beta}^{\dagger}}{i} \right )^2 = N_2 + N - M - 4 |\langle \hat{\alpha}^{\dagger}\hat{\beta}
\rangle |^2 \eea where $N_2 = 2 \langle \hat{\alpha}^{\dagger}\hat{\alpha}  \hat{\beta}^{\dagger}\hat{\beta} \rangle$, $N =
\langle \hat{\alpha}^{\dagger}\hat{\alpha} + \hat{\beta}\hat{\beta}^{\dagger} \rangle$ and \bea M = \langle \hat{\alpha}^{\dagger
2} \hat{\beta}^2 \rangle  + \langle  \hat{\alpha}^{ 2} \hat{\beta}^{\dagger 2 } \rangle  - \langle
\hat{\alpha}^{\dagger}\hat{\beta} + \hat{\alpha}\hat{\beta}^{\dagger} \rangle^2  \eea Collecting all these terms, the
inequality (\ref{ineq1})  can be expressed in the form \bea  (N_2 +  N + M) (N_2 + N - M - 4 |\langle
\hat{\alpha}^{\dagger}\hat{\beta} \rangle |^2) \ge |N|^2 \label{ineq} \eea Violation of this inequality means the
occurrence of entanglement. It is worth mentioning here that all these criteria are sufficient for showing entanglement in a bipartite system.

We now concentrate on the form of the state given in Eq. (\ref{newstate}). In all the joint base
states $\mid m, n\rangle$, quasiparticle vacuum state  ($\mid 0 \rangle_S$)   of either condensate $S$ (A ,
B)  appears explicitly. This means that the correlation function $N_2$ of the quasiparticle number operators
is zero for any parameter regime. This reduces the inequality (\ref{ineq}) to the form \bea  - M^2 - 4 ( N +
M)  |\langle \hat{\alpha}^{\dagger}\hat{\beta} \rangle |^2) \ge 0  \label{ineqquasi} \eea Since $N_2 = 0$, the
variance $\Delta \left ( \hat{\alpha}^{\dagger} \hat{\beta}+ \hat{\alpha}\hat{\beta}^{\dagger} \right )^2$ equals to  $ ( N +
M)$ . Because this variance is non-negative, the quantity   $ ( N + M)$ must be non-negative.  This
implies that the inequality (\ref{ineqquasi}) is violated. Thus we have proved that the state (17) is entangled.
 \begin{figure}
 \includegraphics[width=4.125in]{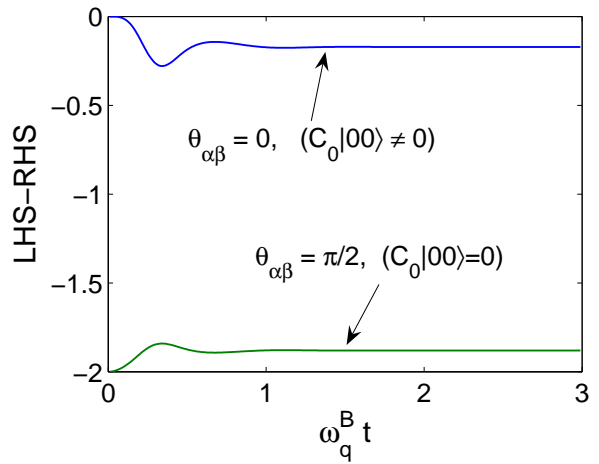}
 \caption{The effect of Gaussian component of the entangled state (17) on the violation of
entanglement inequality (\ref{ineq}) in the quasiparticle picture is illustrated. The upper curve corresponds
to the case when the purely Gaussian basis state $\mid 00 \rangle$ is present in (17), and the lower curve
corresponds to the case when this state is eliminated by choosing the appropriate  phases of the
laser fields. For two identical
 BEC's and same atom-field couplings, when the phase difference between the two input probe  beams is
 $\pi/2$, the coefficient
of the basis $\mid 00 \rangle$ vanishes provided the path-difference between the two output probe beams
arriving at the detectors is zero. Here $n_p = 10$ and all other parameters are same as in Fig.2.  }
 \label{figdel}
 \end{figure}
\subsection{Stronger entanglement by selection of probe phase}
Here we show that the entanglement can be made stronger by choosing appropriate phase
of the two probe beams.
In state (\ref{newstate}), there is a purely Gaussian component corresponding to the basis $\mid 0,0\rangle$
which implies that both condensates are in their respective quasiparticle vacuum.  Since the criterion
(\ref{ineq}) is devised for verifying entanglement in  non-Gaussian bipartite state, we expect that, on
elimination of this purely Gaussian component from  (\ref{newstate}), violation of the inequality
(\ref{ineq})  should be much stronger.  This component can be eliminated if the coefficient $C_0$ given in
Eq. (\ref{c0})  is made to vanish. For simplicity, let us assume that both the condensates are identical and
their atom-field coupling constants are the same.  This means $a_c = b_c$. Furthermore, we assume that
$\Delta \phi  = 0$. In such a situation, to make $C_0$ vanish implies $\alpha^2 = - \beta^2$ where $\alpha$
and $\beta$ are the amplitude of the two input probe fields which are assumed to be in the coherent states.
In other words, this means $\alpha = i \beta$ that is the two input probes should have a phase difference
$\theta_{\alpha\beta}$ equal to  $\pi/2$. Now, a phase difference of $\pi/2$ between two  beams can easily
be made if they are derived from a single laser through a  50:50 beam splitter with transmitted and
reflected beams being  used  as the two input beams. We plot left hand side (LHS) minus right hand side
(RHS) of Eq. (\ref{ineq}) as a function of interaction time in Figs. 3 and 4 for various parameters.
Obviously, negativity of (LHS-RHS) implies violation of the inequality   (\ref{ineqquasi}) and hence
entanglement. For all our numerical illustrations, we consider two identically prepared BEC's with the same
atom-field coupling strength. However, it is worth pointing out that this assumption is not necessary for
exploration of entanglement between the condensates, this is assumed only for the sake of simplicity.
Figure 3 shows the effect of probe photon number (or intensity) on the degree of violation of the
inequality. Weaker the probe intensity is, stronger is the violation. Figure 4 illustrates that when the
purely Gaussian component is eliminated ($\theta_{\alpha\beta} = \pi/2$)  from the  state (17),
the negativity of (LHS-RHS) becomes much stronger implying that the criterion (\ref{ineq}) works better for
highly non-Gaussian entangled states. For identical coupling constants and detuning between the pump and the
probe for the two condensates, we find that the degree of violation of the inequality saturates at the long
interaction time limit. However, for any mismatch in coupling constants or detunings, we have found that
(LHS-RHS) tends to zero in the long interaction time regime (not shown). The effect of atom-field coupling
strength on the  negativity of (LHS-RHS) is illustrated in Fig. 5. We notice that in the strong-coupling
regime the negativity is stronger. Furthermore, in the limit of very strong-coupling the violation of the
inequality becomes almost insensitive to the initial phase-difference $\theta_{\alpha\beta}$ of the two
probe beams. This indicates that strong-coupling is important for entanglement between remote condensates.
This in turn brings in the important role  of condensates in quantum information science: Since the
effective atom-field coupling strength is proportional to the square root of number of atoms,
strong-coupling regime can easily be attained with a BEC rather than a thermal gas. Since thermal and phase
fluctuations in a BEC is at the minimum level, collective atom-field coupling can easily be accomplished
with a BEC in a cavity with a moderate Q-factor. For experimental verification of entanglement, one needs to  measure the various
variances (appearing in the inequality) in the phonon or collective excitation modes of BEC.

\begin{figure}
 \includegraphics[width=4.125in]{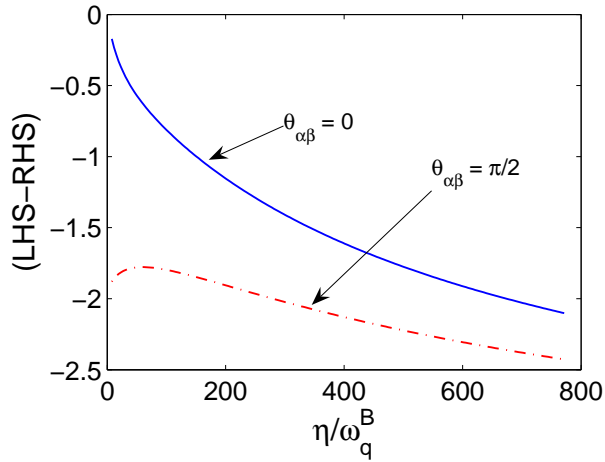}
 \caption{Violation of the inequality (\ref{ineq}) as a function of coupling constant $\eta_A = \eta_B = \eta$
  for
two cases $\theta_{\alpha\beta} = 0$ (solid) and  $\theta_{\alpha\beta} = 0$ (dashed) which correspond to
presence and absence of the purely Gaussian basis $\mid 0 0 \rangle $ in the entangled state. Here  $\omega_q^B t = 5 $ and $n_p = 10$.}
 \end{figure}

\section{conclusion}
In conclusion, we have shown that two remote independent condensates can be made entangled in collective
excitations of BEC such as quasiparticle or phonon variables by projective measurement on two photons
Bragg-scattered by the two condensates. The generated entangled state is explicitly non-Gaussian and the
existing criterion for entanglement in Gaussian variables is insufficient in revealing entanglement in this
state. This has prompted us to test another  criterion \cite{gsabiswas} introduced  specially for testing entanglement
in non-Gaussian states. We have shown that with this criterion the projective state of the two condensates
is entangled and the entanglement can be made stronger by choosing probe phase in such a way so that the non-Gaussian nature of the entangled state becomes more prominent.  The entanglement is shown to be stronger
when the probe beams are weaker and the atom-field coupling is higher. Since atoms behave collectively in
BEC-field interaction and also because of $\sqrt{N}$ (N is the atom number) scaling of coupling strength,
BEC can play an important role in generation and manipulation of entanglement.  Since the entanglement in
the present context arises in the higher order fluctuations of the paring operators, experimental detection
of this entanglement requires new techniques that will enable to measure the pairing fluctuations of phonons
or quasiparticles. Using Bragg spectroscopy phonons in a BEC have been detected by Ketterle's group. But how to
detect phonon-phonon correlation which is crucially required for exploration of higher order entanglement is
presently unknown. Perhaps by knocking out pairs of atoms from the two trapped condensates and measuring their
fluctuations, one can detect this entanglement.
 Our proposed scheme can be
easily generalized for multiple condensates. Furthermore, it may be interesting to generate entanglement
both in spin and motional degrees of freedom using spinor condensates and polarized lights in our scheme
with  the possibility of interesting interplay  of entanglement in spin and center-of-mass degrees of
freedom.

\begin{center}
{\bf ACKNOWLEDGMENT}
\end{center}
This work is supported by NSF Grant No. 0653494. One of us (BD) gratefully acknowledges the hospitality at
Oklahoma State University where this work was initiated during his visit.

\appendix
\section*{Appendix}
Here we derive the effective hamiltonian for Bragg scattering in a condensate. The total hamiltonian of a
condensate interacting with two single-mode light fields is
 $H=H_{A}+H_{F}+H_{AF}$, where
$H_{F} = \hbar\omega_{1} \hat{c}_{\mathbf{k}_1}^{\dagger}\hat{c}_{\mathbf{k}_1} + \hbar\omega_{2}
\hat{c}_{\mathbf{k}_2}^{\dagger}\hat{c}_{\mathbf{k}_2}$ corresponds to the two fields described by the
operators $\hat{c}_{\mathbf{k}_1}$ and $\hat{c}_{\mathbf{k}_2}$ with photon momentum $\mathbf{k}_1$ and
$\mathbf{k}_2$ and the frequencies $\omega_1$ and $\omega_2$, respectively. We assume $\omega_1 > \omega_2$.
The free part of the atomic hamiltonian
\begin{eqnarray}
&& H_{A} =\sum_{k}\hbar\omega_{k} \hat{\pi}_{{\mathbf{k}}}^{\dagger}\hat{\pi}_{{\mathbf{k}}} +
\frac{4\pi\hbar^2a_s}{2mV} \nonumber \\ &\times& \sum_{\mathbf{k}_3,\mathbf{k}_4,\mathbf{k}_5,\mathbf{k}_6}
\hat{\pi}_{\mathbf{k}_3}^{\dagger}\hat{\pi}_{\mathbf{k}_4}^{\dagger}
\hat{\pi}_{\mathbf{k}_5}\hat{\pi}_{\mathbf{k}_6}
\delta_{\mathbf{k}_{3}+\mathbf{k}_4,\mathbf{k}_5+\mathbf{k}_6}
\end{eqnarray}  governs the dynamics of
a weakly interacting atomic condensate and
\begin{eqnarray}
 H_{AF} &=&\hbar\Omega
\hat{c}_{\mathbf{k}_2}^{\dagger}\hat{c}_{\mathbf{k}_1}
\sum_{k}\left(\hat{\pi}_{\mathbf{q+k}}^{\dagger}\hat{\pi}_{{\mathbf{k}}}
+\hat{\pi}_{-\mathbf{q}+\mathbf{k}}\hat{\pi}_{{\mathbf{k}}}^{\dagger} \right)\nonumber \\ & +& {\rm H.c.}
\label{hI}
\end{eqnarray}
describes atom-field  interaction. Here $\hat{\pi}_{k}(\hat{\pi}_{k}^{\dagger})$ is the
annihilation(creation) operator of an  atom with momentum $\mathbf{k}$ and  frequency
$\omega_{k}=\frac{\hbar k^2}{2m}$; ${\mathbf{q}} = {\mathbf{k}}_{1} - {\mathbf{k}}_{2}$,
$\Omega=(\vec{E}_{1}.\vec{d}_{13}) (\vec{E}_{2}.\vec{d}_{32})/(\hbar^2\Delta)$ is the two-photon Rabi
frequency, where $E_{1(2)}$ are the  field amplitudes,  the $\vec{d}_{ij}$ is the electronic transition
dipole moment between the states $|i\rangle$ and $|j\rangle$ of an atom and $\Delta$ is the detuning of the
first laser field (with frequency $\omega_1$)
 from the transition frequency between the electronic ground ($|1\rangle$)
and excited ($|3\rangle$) levels of the atom. For a single-component condensate the electronic ground states
$|1\rangle$ and $|2\rangle$  are the same. Here $a_s$ is the s-wave scattering length of the atoms and $V$
is the volume of the condensate.

 Using Bogoliubov's prescription
$\hat{\pi}_{0},\hat{\pi}_{0}^{\dagger} \rightarrow \sqrt{N_{0}}$, and keeping the number density
$n_{0}=N_{0}/V$ fixed in the thermodynamic limit, one can transform  the hamiltonian $H_{A}$ into a
quadratic form. Further, applying  Bogoliubov's transformation it is possible to  diagonalize $H_{A}$ and
rewrite the entire hamiltonian in terms of Bogoliubov's  quasi-particle operators $\hat{\chi}_{\mathbf{k}}$.
Considering the condensate ground state energy as the zero of the energy scale, and treating the laser
light with higher frequency ($\omega_1$) classically,  the effective hamiltonian can be written as
\begin{eqnarray}
H_{eff} &=& \hbar \omega_{q}^{B}\left(\hat{\chi}_{{\mathbf{q}}}^{\dagger} \hat{\chi}_{{\mathbf{q}}}
+\hat{\chi}_{-{\mathbf{q}}}^{\dagger} \hat{\chi}_{-{\mathbf{q}}}\right)
-\hbar\delta\hat{c}_{\mathbf{k}_2}^{\dagger} \hat{c}_{\mathbf{k}_2} \nonumber \\ &+&\left[\hbar\eta
\hat{c}_{\mathbf{k}_{2}}^{\dagger} (\hat{\chi}_{\mathbf{q}}^{\dagger} + \hat{\chi}_{-{\mathbf{q}}})
+{\mathrm H.c.}\right]
\end{eqnarray}
where $\delta=\omega_{1}-\omega_{2}$ and $\eta = \sqrt{N}f_q\Omega$; where  $f_q = u_q-v_q$. In writing the
above equation, we have retained only two dominant momentum side-modes of the condensate under the
assumption of Bragg resonance ($\delta \simeq \omega_{q}$).

\end{document}